\documentclass[useAMS,usenatbib]{mn2e}
\usepackage{graphicx,amssymb,color}
\usepackage[normalem]{ulem}

\newcommand{\rev}{  }

\title[Tides between giant planets and white dwarfs]
{Tidal circularization of gaseous planets orbiting white dwarfs}

\author[]{Dimitri Veras$^{1,2}$\thanks{E-mail: d.veras@warwick.ac.uk}\thanks{STFC Ernest Rutherford Fellow},
Jim Fuller$^3$
\\
$^{1}$Centre for Exoplanets and Habitability, University of Warwick, Coventry CV4 7AL, UK
\\
$^{2}$Department of Physics, University of Warwick, Coventry CV4 7AL, UK
\\
$^{3}$TAPIR, Mailcode 350-17, California Institute of Technology, Pasadena, CA 91125, USA, USA
}

\pubyear{2019}

\begin{document}
\label{firstpage}
\pagerange{\pageref{firstpage}--\pageref{lastpage}}
\maketitle

\begin{abstract}
A gas giant planet which survives the giant branch stages of evolution at a distance of many au and then is subsequently perturbed sufficiently close to a white dwarf will experience orbital shrinkage and circularization due to star-planet tides. The circularization timescale, when combined with a known white dwarf cooling age, can place coupled constraints on the scattering epoch as well as the active tidal mechanisms. Here, we explore this coupling across the entire plausible parameter phase space by computing orbit shrinkage and potential self-disruption due to chaotic f-mode excitation and heating in planets on orbits with eccentricities near unity, followed by weakly dissipative equilibrium tides. We find that chaotic f-mode evolution activates only for orbital pericentres which are within twice the white dwarf Roche radius, and {\rev easily restructures or destroys ice giants but not gas giants. This type of internal thermal destruction provides an additional potential source of white dwarf metal pollution.} Subsequent tidal evolution {\rev for the surviving planets} is dominated by non-chaotic equilibrium and dynamical tides which may be well-constrained by observations of giant planets around white dwarfs at early cooling ages.
\end{abstract}


\section{Introduction}

The recent discovery of a planetesimal orbiting a white dwarf well within its Roche radius for strengthless rubble piles suggests that this minor planet is actually a ferrous fragment of a core of a major planet \citep{manetal2019}. Despite the uniqueness and startling nature of this find, in fact such a configuration is consistent with theoretical constructs about the fate of major planets \citep{veras2016a}.

In the solar system, at least five major planets -- including the four giants -- will survive the Sun's giant branch phases of evolution \citep{schcon2008,veras2016b}. Subsequent evolution of Jupiter, Saturn, Uranus and Neptune is quiescent, but only by dint of fortuitous mutual spacing which avoids resonances and is not quite small enough to trigger instability \citep{dunlis1998,debsig2002,veretal2013a,voyetal2013}.

Alternatively, a planetary system like HR 8799, which contains four gas giant planets on more tightly packed and resonant orbits \citep{maretal2008,maretal2010,gozmig2014,wanetal2018}, may experience a very different fate. Several investigations reveal that packed planetary systems of three or more planets around single stars can survive the entire main sequence and giant branch phases, only to experience at least one instance of gravitational scattering during the white dwarf phase \citep{musetal2014,vergae2015,veretal2016,musetal2018}. In fact multi-planet systems are not even necessary to incite gravitational instability during the white dwarf phase, as a binary stellar companion could also accomplish the same task \citep{bonver2015, hampor2016,petmun2017,steetal2017,veretal2017a,steetal2018}.

One potential outcome of such gravitational instability is a kick that places a planet on a highly eccentric ($e > 0.99$) orbit \citep{caretal2019}. Many investigators have quantified the rate at which minor planets such as asteroids or comets that are kicked on highly eccentric orbits accrete onto the white dwarf \citep{alcetal1986,bonetal2011,debetal2012,frehan2014,veretal2014,stoetal2015,caihey2017,musetal2018,smaetal2018,smamar2019} or approach within the vicinity of its Roche radius  \citep{veretal2015a,broetal2017}. A strong motivation for these studies has been an understanding of the planetary debris seen in the atmospheres of over 1000 white dwarfs \citep{kleetal2013,kepetal2015,kepetal2016,holetal2017,holetal2018,haretal2018}, particularly as the entire known population of white dwarfs has increased by an order of magnitude in the year 2018 \citep{genetal2019}. Another strong motivation is understanding the dynamical history of the asteroid which is currently orbiting and disintegrating around the white dwarf WD 1145+017 \citep{vanetal2015} on a near-circular orbit \citep{guretal2017,veretal2017b}.

The fate of major planets on highly eccentric orbits which approach a white dwarf has not been modelled in nearly as much detail, {\rev partly because such planets have not yet been found. Few white dwarfs have been observed well enough to detect transits, and radial velocity techniques are ineffective at detecting non-transiting planets orbiting white dwarfs. Nevertheless, many investigators have previously attempted to detect major planets orbiting white dwarfs with a variety of methods \citep{buretal2002,hogetal2009,debetal2011,faeetal2011,steetal2011,fuletal2014,xuetal2015,sanetal2016,rowetal2019}. 

However, the {\it K2} mission ushered in a new era of discovery. WD 1145+017 was first seen by {\it K2} \citep{vanetal2015}, prompting \cite{vanvan2018} to compute {\it K2} white dwarf planet occurence rates through transit photometry as a function of mass and distance. They found a strong dependence on both parameters, and their Figs. 2-3 illustrate that the occurence rate can vary by tens of per cent within the regime where tides may be active. Now, other missions such as TESS, LSST \citep{lunetal2018,corkip2018} and Gaia \citep{peretal2014} will provide additional opportunities. In particular, the last data release for Gaia is expected to detect about one dozen giant planets orbiting white dwarfs through astrometry \citep{peretal2014}.

Despite these promising prospects,} there is a dearth of studies investigating the mechanical destruction of a planet entering a white dwarf's Roche radius. Dedicated investigations of planet-white dwarf tidal interactions are limited to solid planets without surface oceans \citep{veretal2019,verwol2019}. Solid body tidal mechanisms cannot be applied to gas giant planets, which require a completely different treatment. Because white dwarfs are negligibly tidally distorted by planetary companions, tidal interaction mechanisms between a white dwarf and other stars \citep{fullai2011,fullai2012,fullai2013,fullai2014,valetal2012,sraetal2014,vicetal2017,mcnetal2019} are also not necessarily suitable. However, other stars, through their fluid-like nature, do have stronger links to giant planets.

In this paper, we model the tidal interaction between a gas giant planet and a white dwarf. This interaction may be split into two regimes: a high-eccentricity {\rev regime ($e \gtrsim 0.95$) where the motion may be dominated by chaotic energy exchange between internal modes and angular orbital momentum \citep{mardling1995a,mardling1995b,ivapap2004,ivapap2007,viclai2018,wu2018,tesetal2019,vicetal2019}}, and a post-chaos regime where orbit shrinkage and circularization are dominated by equilibrium tides \citep{alexander1973,hut1981}.

A beneficial feature of white dwarfs is that their observable properties allow us to estimate their ``cooling age'', or the time since they were born, typically to much better accuracy than the age of a main sequence star.  Assume that a giant planet underwent a gravitational instability at a time $t_{\rm sca}$ after the white dwarf was born, and sometime later is observed on a near-circular orbit just outside the Roche radius of a white dwarf with a cooling age of $t_{\rm cool}$. The planet might have experienced the chaotic tidal regime first for a time interval of $\tau_{\rm chaos}$, which could equal zero. Immediately afterwards it experienced the non-chaotic tidal regime for a time interval of $\tau_{\rm non-chaos}$, until the planet's orbit circularized. Then

\begin{equation}
\underbrace{t_{\rm cool}}_{\rm observed} 
> 
\underbrace{t_{\rm sca}}_{\rm unknown} 
+ 
\underbrace{\tau_{\rm chaos}}_{\rm computed \ here} 
+ 
\underbrace{\tau_{\rm non-chaos}}_{\rm estimated \ here}
.
\label{master}
\end{equation}

Equation (\ref{master}) suggests that a combination of observations and theory can constrain $t_{\rm sca}$, which in turn helps us trace the dynamical history of a given planetary system. Our focus here is to compute $\tau_{\rm chaos}$ across the entire available phase space for white dwarf planetary systems by specifically using the iterative map as presented in \cite{vicetal2019}  (Section 2), and then to estimate $\tau_{\rm non-chaos}$ by using a simplified prescription for tidal quality functions (Section 3). We discuss our results in Section 4, and conclude in Section 5. Table \ref{description} provides a helpful chart of every variable used in this paper; we took care to maintain consistency with the notation used in \cite{vicetal2019} for easy reference.

\begin{table*}
 \centering
 \begin{minipage}{180mm}
 \centering
  \caption{Variables used in this paper, with Roman variables first followed by the Greek ones.
Quantities with overhead tildas, which are not shown here, are scaled according to $x = \tilde{x} \sqrt{G M_{\rm p}/ R_{\rm p}^3}$ and $\tilde{x} = x \sqrt{R_{\rm p}^3 / (G M_{\rm p})}$.}
  \label{description}
  \begin{tabular}{@{}llll@{}}
  \hline
  Variable & Explanation & Units & Equation \\
  \hline
$a$ & Semimajor axis of orbit & Length & \ref{akeq}, \ref{dadtequil} \\[2pt]
$c_{\alpha}$ & The dominant f-mode (includes amplitude and phase) & Angle/Time & \ref{Etoohigh} \\[2pt]
$\Delta c_{\alpha}$ & Change in dominant f-mode amplitude from pericentre passage & Angle/Time & \ref{delceq} \\[2pt]
$e$ & Eccentricity of orbit & Dimensionless & \ref{ekeq}, \ref{dedtequil} \\[2pt]
$E_{\alpha}$ & Energy of dominant f-mode & Mass $\times$ Length$^2$ / Time$^2$ & \ref{deltEalpeq} \\[2pt]
$\Delta E_{\alpha}$ & Change in energy of dominant f-mode amplitude from pericentre passage & Mass $\times$ Length$^2$ / Time$^2$ & \ref{delEeq} \\[2pt]
$E_{\rm B}$ & Energy of orbit & Mass $\times$ Length$^2$ / Time$^2$ & \ref{Ebkeq} \\[2pt]
$E_{\rm bind}$ & Binding energy of planet & Mass $\times$ Length$^2$ / Time$^2$ & \ref{Ebind} \\[2pt]
$E_{\rm max}$ & {\rev Maximum energy before non-linear effects become important} & Mass $\times$ Length$^2$ / Time$^2$ & \ref{Emaxeq} \\[2pt] 
$E_{\rm resid}$ & {\rev Residual energy after a thermalisation} & Mass $\times$ Length$^2$ / Time$^2$ & \ref{Eresid} \\[2pt] 
$f$ & Functions of eccentricity from Hut (1981) & Dimensionless & \ref{fstart}-\ref{fend} \\[2pt]
$k$ & Counter for number of pericentre passages & Dimensionless &  \\[2pt]
$K_{22}$ & Hansen coefficient & Dimensionless & \ref{k22eq} \\[2pt]
$M_{\ast}$ & Mass of white dwarf & Mass & \\[2pt]
$M_{\rm p}$ & Mass of (giant) planet & Mass & \\[2pt]
$P$ & Orbital period & Time & \ref{orbeq} \\[2pt]
$Q_{\ast}'$ & Modified white dwarf tidal quality factor & Dimensionless & \\[2pt]
$Q_{\rm p}'$ & Modified planetary tidal quality factor & Dimensionless & \\[2pt]
$Q_{\alpha}$ & Tidal overlap integral & Dimensionless & \\[2pt]
$r_{\rm p}$ & Orbital pericentre & Distance & \ref{perieq} \\[2pt]
$r_{\rm Roche}$ & Roche radius of the white dwarf for a spinning fluid planet & Distance & \ref{roche} \\[2pt]
$R_{\ast}$ & Radius of white dwarf & Length & \\[2pt]
$R_{\rm p}$ & Radius of (giant) planet & Length & \\[2pt]
$S_{\ast}$ & Spin rate of the white dwarf & Angle/Time & \\[2pt]
$t_{\rm cool}$ & Time since the white dwarf was born (the ``cooling age'') & Time & \ref{master} \\[2pt]
$t_{\rm sca}$ & Time of gravitational scattering since white dwarf was born & Time & \ref{master} \\[2pt]
$T$ & Auxiliary variable & Dimensionless & \ref{teq} \\[2pt]
$u$ & Multiple of white dwarf Roche radius which equals initial orbital pericentre & Dimensionless & \ref{eu} \\[2pt]
$z$ & Auxiliary variable & Dimensionless &  \ref{zeq} \\[2pt]
$\alpha$ & Mode index & Dimensionless & \\[2pt]
$\gamma$ & Polytropic index for giant planet & Dimensionless & \\[2pt]
$\epsilon_{\alpha}$ & A mode frequency & Angle/Time & \ref{epseq} \\[2pt] 
$\eta$ & Auxiliary variable & Dimensionless & \ref{etaeq} \\[2pt]
$\rho_{\rm p}$ & Density of planet & Mass/Length$^3$  & \\[2pt]
$\sigma_{\alpha}$ & A mode frequency & Angle/Time & \ref{sigeq} \\[2pt] 
$\tau_{\rm chaos}$ & Timescale over which chaotic f-mode evolution dictates evolution & Time & \ref{master} \\[2pt]
$\tau_{\rm chaos, ana}$ & {\rev Analytic estimate of $\tau_{\rm chaos}$}  & Time & \ref{chaosana} \\[2pt]
$\tau_{\rm non-chaos}$ & Timescale from the end of chaotic f-mode evolution to circularization  & Time & \ref{master}, \ref{empirical} \\[2pt]
$\omega_{\alpha}$ & A mode frequency & Angle/Time & \ref{omeq} \\[2pt] 
$\Omega_{\rm p}$ & Orbital frequency at the orbital pericentre & Angle/Time & \ref{ompeq} \\[2pt]
$\Omega_{\rm s}$ & ``Pseudosynchronous'' spin rate of the planet & Angle/Time & \ref{omseq} \\[2pt]
\hline
\end{tabular}
\end{minipage}
\end{table*}

\section{Chaotic tidal regime}

In this section we determine $\tau_{\rm chaos}$, the timescale over which the giant planet's orbital evolution is dominated by the chaotic excitation of internal modes. We follow the iterative map procedure in \cite{vicetal2019}, but scaled to the architecture of a giant planet orbiting a white dwarf (with mass $M_{\ast} = 0.6M_{\odot}$, a value we adopt throughout the paper). We also apply the procedure across the entire relevant phase space for white dwarf planetary systems, and with a more algorithmic approach; their paper contains more details of the physics and subtleties of the iterative map relations.

\subsection{Single mode evolution}

Our first approximation is that we consider the evolution of one mode only --- the f=2 mode --- within a spinning fluid giant planet that is constructed from an equation of state with polytropic index $\gamma = 2$. Figure 1 of \cite{vicetal2019} illustrates that this unimodal approximation holds for the entire relevant range of orbital pericentres around white dwarfs because the Roche radius of a white dwarf is \citep[Table 1 and Eq. 3 of][]{veretal2017b} 

\begin{equation}
r_{\rm Roche} = 1.619 R_{\odot} \left( \frac{\rho_{\rm p}}{3 \ {\rm g}/{\rm cm}^3} \right)^{-1/3}
\label{roche}
\end{equation}

\noindent{}such that $r_{\rm Roche} = 2.12 R_{\odot} \approx 0.010$~au for a Jupiter-density planet ($\rho_{\rm p} = 1.33$~g/cm$^3$)  and $r_{\rm Roche} = 2.65 R_{\odot} \approx 0.012$~au for a Saturn-density planet ($\rho_{\rm p} = 0.69$~g/cm$^3$). Because our results are sensitive to density, we adopt a generous range of giant planet densities ($0.4 - 17$~g/cm$^3$) by considering $1.0 R_{\rm Jup}$ planets with masses that vary between $0.3 M_{\rm Jup}$ and $13 M_{\rm Jup}$ (spanning the potential range of gas giant planets). 

Given the dependence on density from equation (\ref{roche}), we also do not set a specific initial eccentricity ($e_0$), but rather a pericentre distance $r_{\rm p} = ur_{\rm Roche}$ such that $u>1$.  The initial eccentricity is hence computed from

\begin{equation}
e_{0} = 1 - \frac{u r_{\rm Roche}}{a_0}
.
\label{eu}
\end{equation}

\noindent{}Here $a_0$ is the given initial semimajor axis. One outcome of this study is to determine the relevant range of $u$ and how it varies over the course of an evolution.

The unimodular approximation allows us to establish (from \citealt*{vicetal2019}) the tidal overlap integral $Q_{\alpha} = 0.56$ and obtain the following associated mode frequencies in the rotating frame ($\omega_{\alpha, k-1}$), the inertial frame ($\sigma_{\alpha, k-1}$) and for a non-rotating planet in the slow rotation limit ($\epsilon_{\alpha}$):

\begin{equation}
\epsilon_{\alpha} = 1.22 \sqrt{\frac{G M_{\rm p}}{R_{\rm p}^3}},
\label{epseq}
\end{equation}

\begin{equation}
\sigma_{\alpha, k-1} =  \epsilon_{\alpha} + \Omega_{{\rm s}, k-1},
\label{sigeq}
\end{equation}

\begin{equation}
\omega_{\alpha, k-1} = \epsilon_{\alpha} - \Omega_{{\rm s}, k-1}.
\label{omeq}
\end{equation}

\noindent{}Here, $R_{\rm p}$ is the planet radius, $\alpha$ is the mode index, $k-1$ indicates the number of pericentre passages already experienced since the scattering event,  and $\Omega_{\rm s, k-1}$ is the spin of the planet. Every variable with a subscript of $k-1$ or $k$ must be computed respectively before and after every pericentre passage. One of these variables is the spin of the planet, which is assumed to rotate pseudosynchronously as 

\begin{equation}
\Omega_{\rm s, k-1} = \frac{f_2\left(e_{k-1}\right)}{\left( 1 - e_{k-1}^2 \right)^{3/2}  f_5\left(e_{k-1}\right)}
\sqrt{\frac{G\left(M_{\ast}+M_{\rm p} \right)}{a_{k-1}^3}}
\label{omseq}
\end{equation}

\noindent{}where the $f$ eccentricity functions are from \cite{hut1981}:

\begin{equation}
f_1\left(e\right) = 1 + \frac{31}{2}e^2 + \frac{255}{8}e^4 + \frac{185}{16}e^6 + \frac{25}{64}e^8,
\label{fstart}
\end{equation}

\begin{equation}
f_2\left(e\right) = 1 + \frac{15}{2}e^2 + \frac{45}{8}e^4 + \frac{5}{16}e^6,
\end{equation}

\begin{equation}
f_3\left(e\right) = 1 + \frac{15}{4}e^2 + \frac{15}{8}e^4 + \frac{5}{64}e^6,
\end{equation}

\begin{equation}
f_4\left(e\right) = 1 + \frac{3}{2}e^2 + \frac{1}{8}e^4,
\end{equation}

\begin{equation}
f_5\left(e\right) = 1 + 3e^2 + \frac{3}{8}e^4.
\label{fend}
\end{equation}

\subsection{Criterion for starting chaotic evolution}

Our next consideration is to determine under what conditions chaotic mode evolution {\rev can be initiated}. Not every scattering incident will produce an architecture which is dictated by chaotic evolution, and {\rev we need to identify which do.} The criterion for the initiation of chaotic mode evolution is expressed in Eq. (28) of \cite{vicetal2019}, which we re-write as

\[
1 < \frac{6 \pi \sigma_{\alpha, k-1}}{\left(1-e_{k-1}\right)^{5/2}\Omega_{{\rm p}, k-1}}
      \left( \frac{M_{\rm p}}{M_{\ast}} \right)^{2/3}
      \eta_{k-1}^{-5} T_{k-1}
\]

\begin{equation}
\ \ =\frac{12 \pi^3 \sigma_{\alpha, k-1}^2 Q_{\alpha}^2 K_{22, k-1}^2}
                     {\epsilon_{\alpha} \left(1 - e_{k-1} \right)^6}
                     \left( \frac{M_{\ast}}{M_p} \right) \frac{R_{\rm p}^5}{a_{k-1}^{7/2}\sqrt{G\left(M_{\ast} + M_{\rm p} \right)}}
.
\label{crit}
\end{equation}


Equation (\ref{crit}) reveals nontrivial functional dependences because of both the mode frequencies as well as the following additional variables, starting with the Hansen coefficient $K_{22, k-1}$:

\begin{equation}
K_{22, k-1} \approx \frac{2z_{k-1}^{3/2} \exp{\left( -\frac{2}{3}z_{k-1} \right)}}{\sqrt{15}}
                  \left(1 - \frac{\sqrt{\pi}}{4 \sqrt{z_{k-1}}}  \right) \eta_{k-1}^{3/2},
\label{k22eq}
\end{equation}

\begin{equation}
z_{k-1} = \frac{\sqrt{2} \sigma_{\alpha, k-1}}{\Omega_{{\rm p}, k-1}},
\label{zeq}
\end{equation}

\begin{equation}
\Omega_{{\rm p}, k-1} = \sqrt{\frac{G\left(M_{\ast} + M_{\rm p} \right)}{r_{{\rm p}, k-1}^3}},
\label{ompeq}
\end{equation}

\begin{equation}
r_{{\rm p}, k-1} = a_{k-1}\left( 1 - e_{k-1}\right),
\label{perieq}
\end{equation}

\begin{equation}
\eta_{k-1} = \frac{r_{{\rm p}, k-1}}{R_{\rm p}} \left( \frac{M_p}{M_{\ast}} \right)^{1/3},
\label{etaeq}
\end{equation}

\begin{equation}
T_{k-1} = 2 \pi^2 \left(\frac{\sigma_{\alpha, k-1}}{\epsilon_{\alpha}} \right) Q_{\alpha}^2 K_{22,k-1}^2.
\label{teq}
\end{equation}

In subsection \ref{phase}, we will use equation (\ref{crit}) to determine if chaotic evolution
is activated.

\subsection{Propagating the chaotic evolution}

As already mentioned, in order to evolve the orbit in the chaotic regime, we do not solve differential equations but rather use an
iterative map. \cite{ivapap2004} and \cite{ivapap2007} pioneered the use of iterative maps for chaotic tidal evolution: these maps are algebraic,
usually quicker than solving differential equations, and are iterated after each pericentre passage. 

During each passage, energy is transferred from the dominant f-mode to the orbit. The inputs before each passage
are $a_{k-1}$, $e_{k-1}$, $E_{\rm B, k-1}$ and $c_{\alpha, k-1}$, where the latter two variables respectively represent
the orbital energy and mode. The mode is complex (in the mathematical sense), but is initially set to zero; the final
result is relatively insensitive to this choice. The initial orbital energy is

\begin{equation}
E_{\rm B, 0} = -\frac{G M_{\ast} M_{\rm p}}{2 a_0}
.
\label{EB0}
\end{equation}

\noindent{}The outputs after each passage are $a_{k}$, $e_{k}$, $E_{\rm B, k}$ and $c_{\alpha, k}$.

Completing the iteration requires performing the following computations in sequence:

\begin{equation}
\Delta E_{\alpha} = \frac{G M_{\ast}^2 R_{\rm p}^5}{r_{\rm p, k-1}^6} T_{k-1},
\label{delEeq}
\end{equation}

\begin{equation}
\Delta \tilde{c}_{\alpha} = \sqrt{\frac{\Delta E_{\alpha}}{\left| E_{\rm B, 0} \right|}},
\label{deltilc}
\end{equation}

\begin{equation}
\tilde{c}_{\alpha, k-1} = c_{\alpha, k-1} \sqrt{\frac{R_{\rm p}^3}{G M_{\rm p}}},
\label{tilceq}
\end{equation}

\begin{equation}
\Delta E_{\alpha, k} = \left| E_{\rm B, 0} \right| 
                                    \left( \left| \tilde{c}_{\alpha, k-1} + \Delta \tilde{c}_{\alpha} \right|^2 - \left| \tilde{c}_{\alpha, k-1} \right|^2  \right),
\label{deltEalpeq}
\end{equation}

\begin{equation}
E_{{\rm B}, k} = E_{{\rm} B, k-1} - \Delta E_{\alpha, k}
,
\label{Ebkeq}
\end{equation}

\begin{equation}
a_k = \frac{E_{{\rm B}, k-1}}{E_{{\rm B}, k}} a_{k-1}
,
\label{akeq}
\end{equation}

\begin{equation}
e_{k} = \sqrt{1 - \frac{E_{{\rm B}, k}}{E_{{\rm B}, k-1}}\left(1 - e_{k-1}^2  \right)}
,
\label{ekeq}
\end{equation}

\begin{equation}
\Delta c_{\alpha}  = \Delta \tilde{c}_{\alpha} \sqrt{\frac{G M_{\rm p}}{R_{\rm p}^3}}
.
\label{delceq}
\end{equation}

\begin{figure*}
\includegraphics[width=18cm]{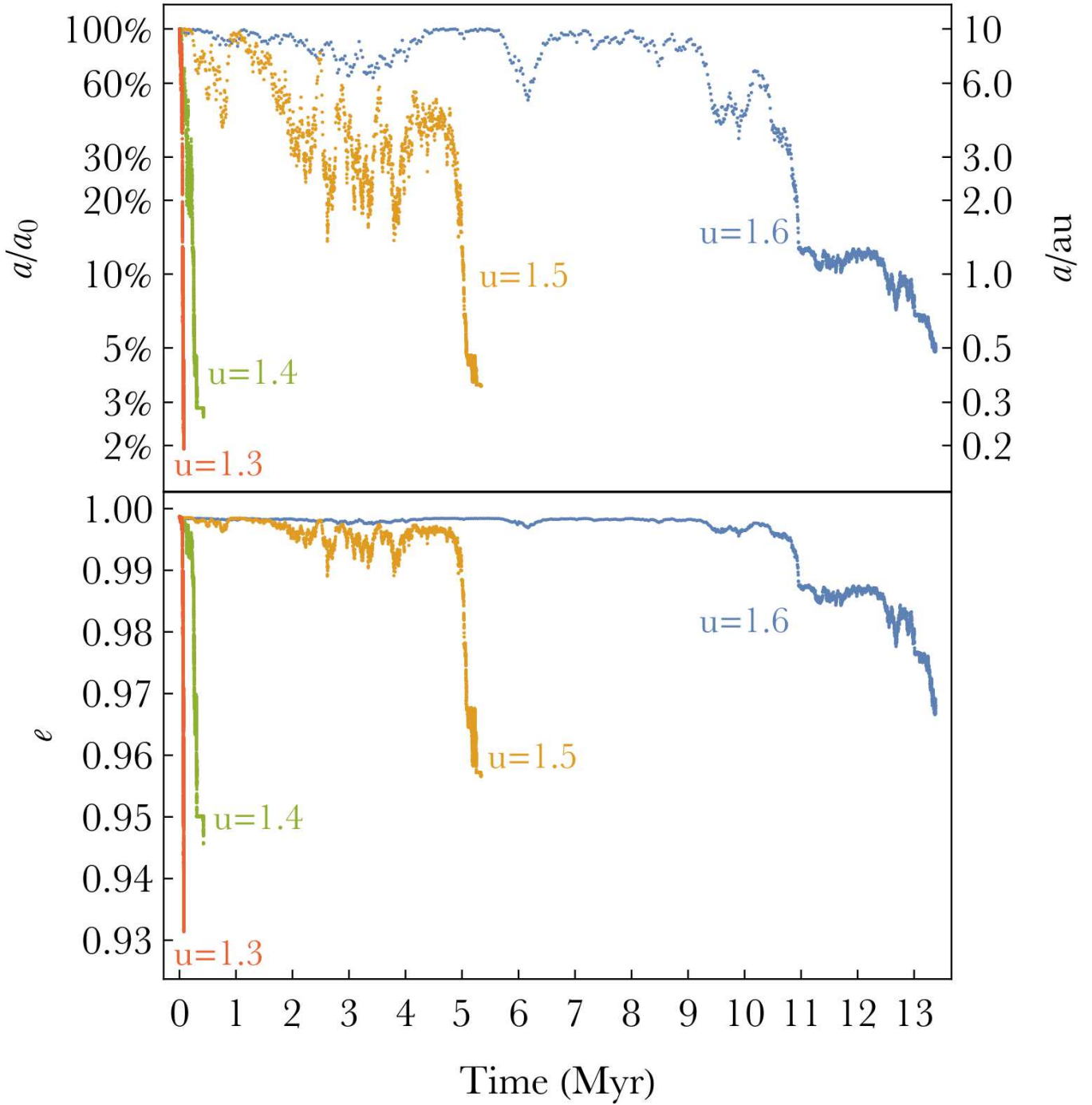}
\caption{
Chaotic orbital evolution of a gas giant planet orbiting a typical $0.6M_{\odot}$ white dwarf solely due to energy exchange with the dominant internal mode of the planet. Only a fraction of the {\rev pericentre passages are plotted as individual points}. The planet properties are $M_{\rm p} = 1M_{\rm Jupiter}$ and $R_{\rm p} = 1R_{\rm Jupiter}$. The initial orbit parameters are what may be expected to be generated from a scattering event which occurred during the white dwarf phase: $a_0 = 10$ au and $u = 1.2-1.6$, such that the orbital pericentre equals $u r_{\rm Roche}$. This chaotic evolution quickly decreases the semimajor axis, and only slightly decreases the eccentricity, before ``turning off'' (equation \ref{critend}) after a time $\tau_{\rm chaos}$ (see equation \ref{master}). Subsequently, because the orbital pericentre is still sufficiently small, non-chaotic tidal effects become dominant.
}
\label{fourcaseJup}
\end{figure*}


\begin{figure*}
\includegraphics[width=8.5cm]{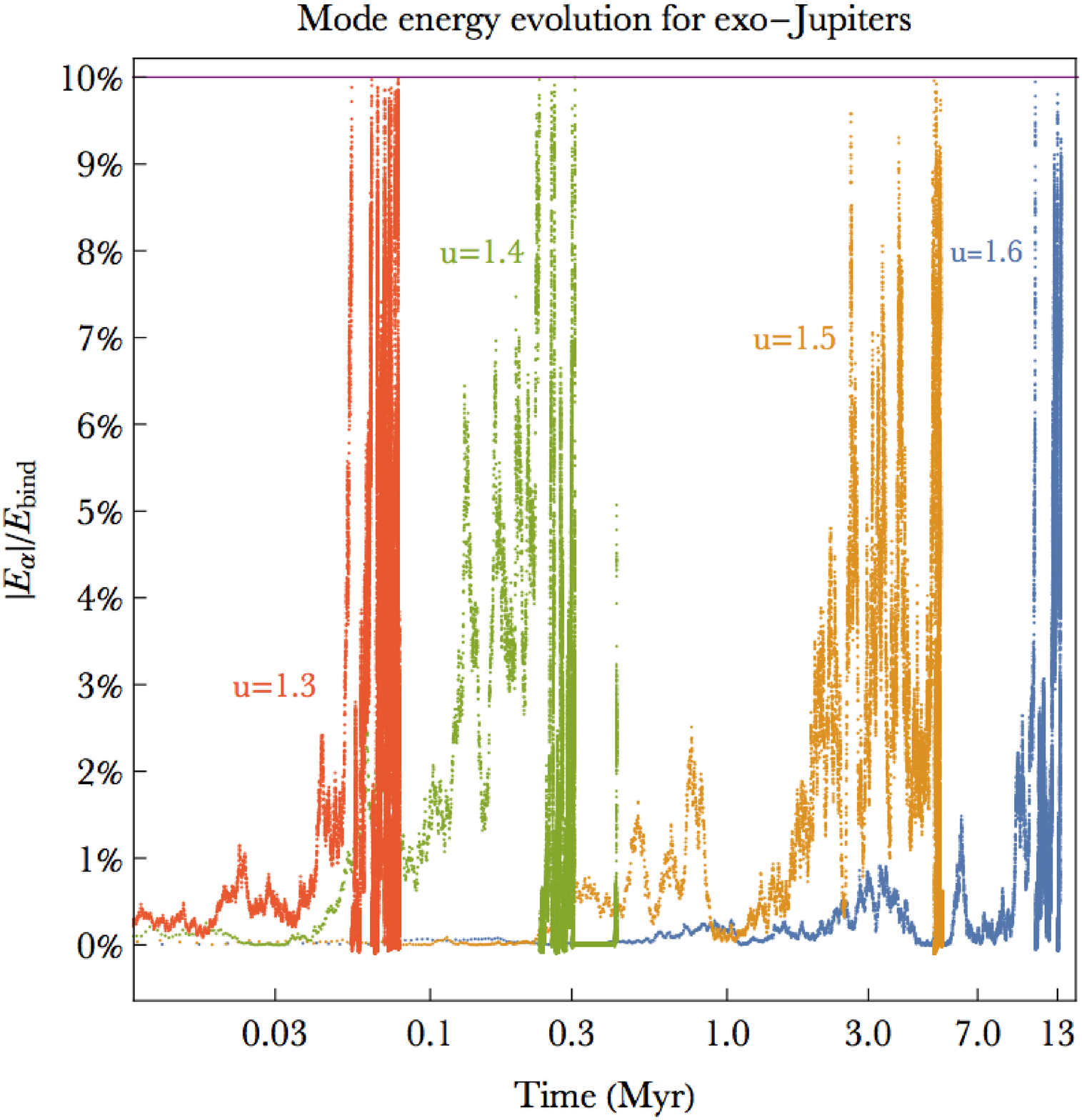}
\includegraphics[width=8.5cm]{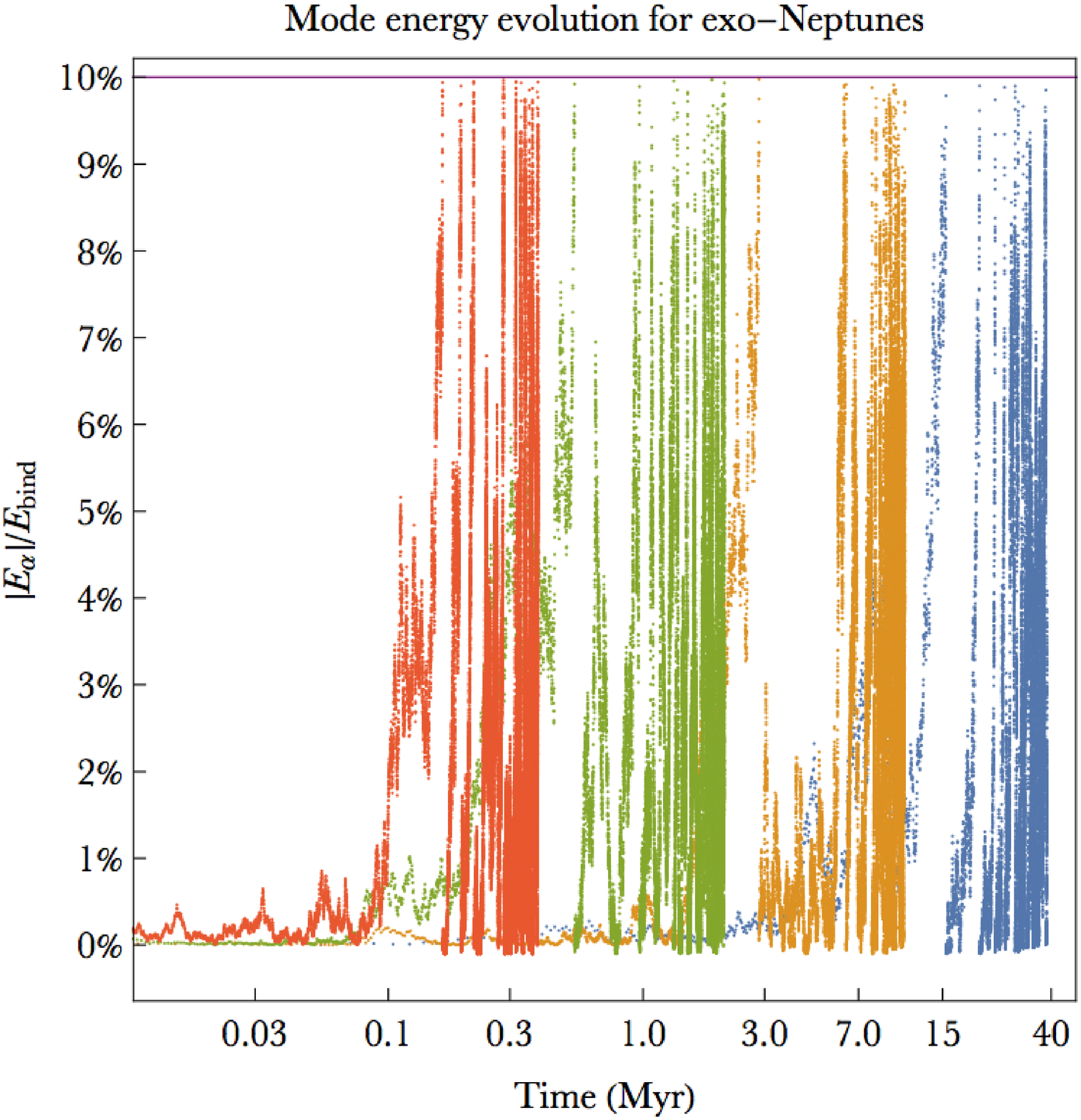}
\caption{
Energy evolution of the planetary f-modes from Fig. \ref{fourcaseJup} relative to their binding energy (left panel). {\rev The right panel illustrates the energy evolution of Neptune-mass and Neptune-radius planets}. As in Vick et al. (2019), we assume that when the mode energy reaches $E_{\rm max} = 0.1E_{\rm bind}$ (equation \ref{Emaxeq} and horizontal purple line on the plots), that energy is thermalized and the mode amplitude is reset (equation \ref{Etoohigh}) due to non-linear effects, which are not modelled here. For the exo-Jupiters, {\rev 7, 5, 4 and 2 thermalization events occur respectively for the $u=1.3, 1.4, 1.5, 1.6$ cases. All of our exo-Neptune models experience more than 10 thermalization events (except for the $u=1.6$ run), at which points the planets may become inflated or disrupted.} Only a fraction of the data points (each corresponding to an individual pericentre passage) is plotted, {\rev except for the $u=1.3$ cases, where every data point is plotted.}
}
\label{energyevo}
\end{figure*}

\begin{figure*}
\includegraphics[width=8.5cm]{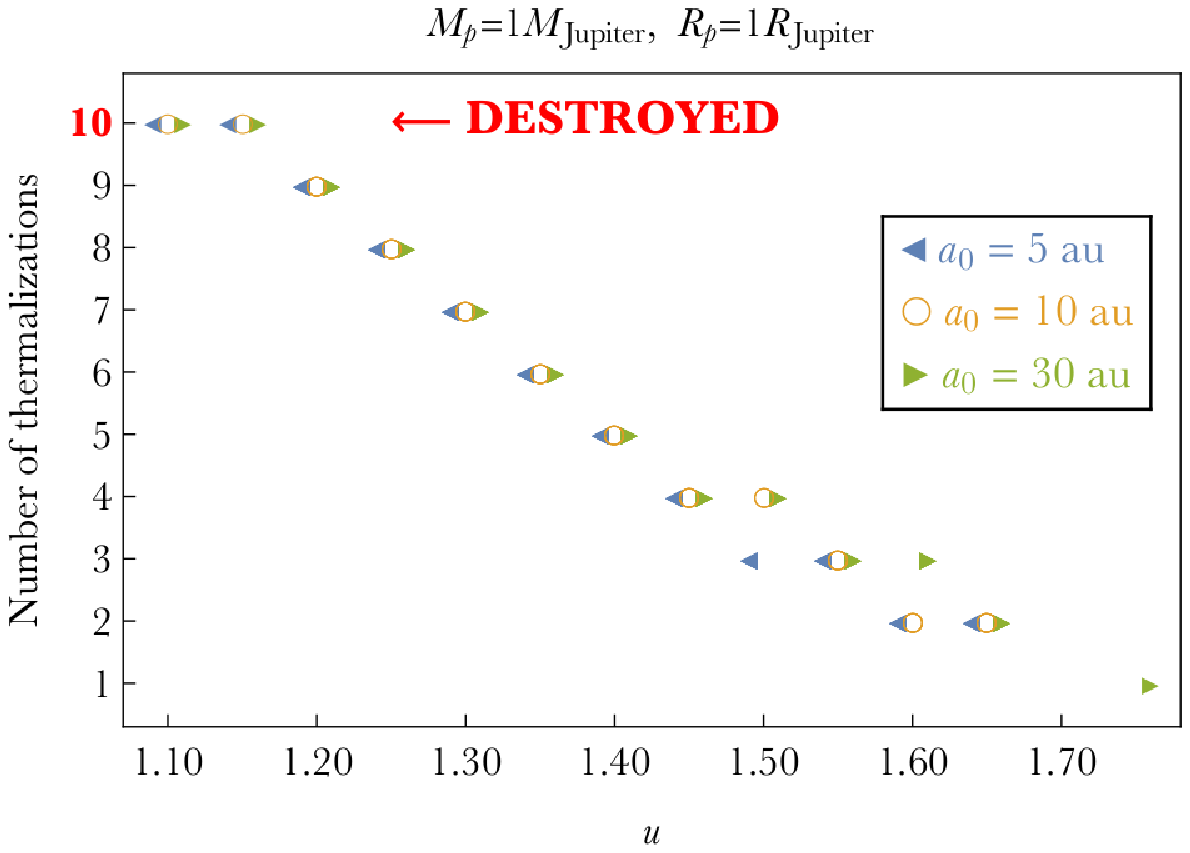}
\includegraphics[width=8.5cm]{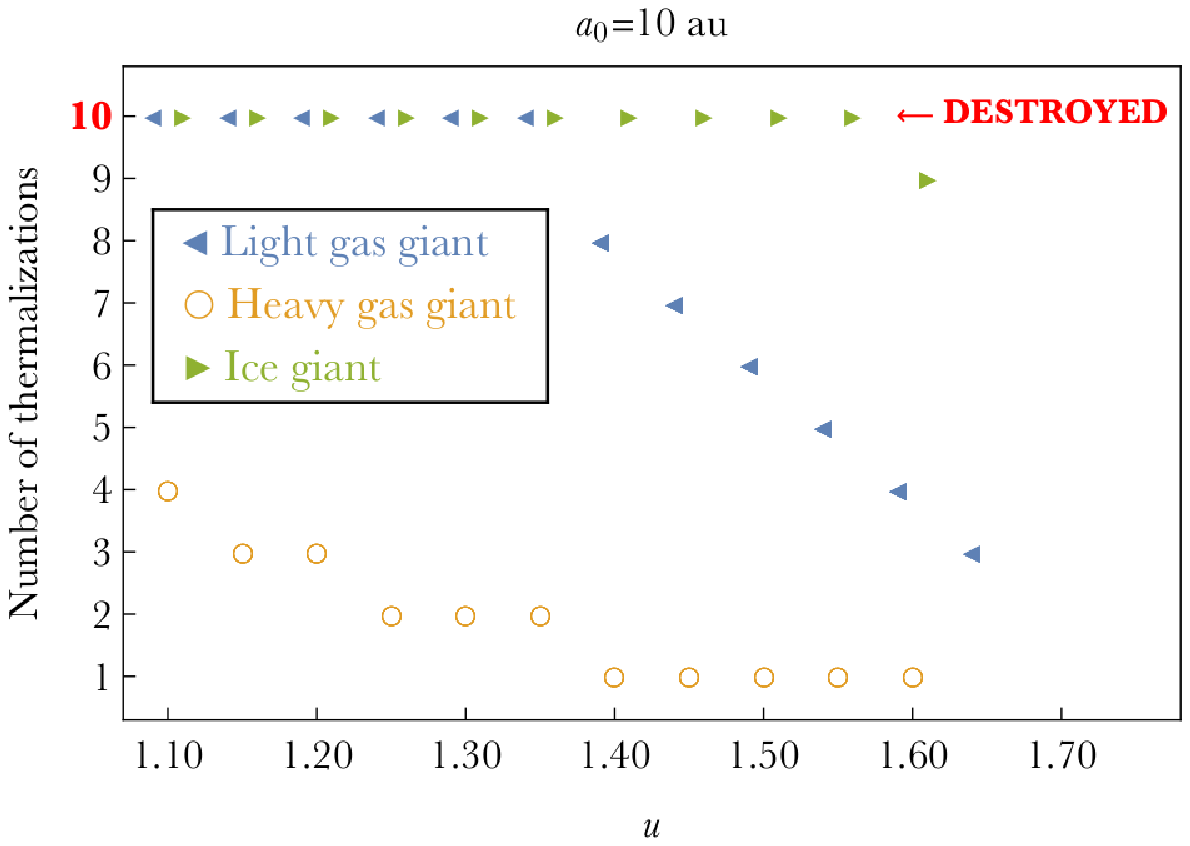}
\caption{
Number of thermalization events across the phase space of initial semimajor axis and physical properties. ``Light Gas Giant'' corresponds to $M_{\rm p} = 0.3 M_{\rm Jupiter}$ and $R_{\rm p} = 1.0R_{\rm Jupiter}$, ``Heavy Gas Giant'' to $M_{\rm p} = 13 M_{\rm Jupiter}$ and $R_{\rm p} = 1.0R_{\rm Jupiter}$, and ``Ice Giant'' to $M_{\rm p} = 1.0 M_{\rm Neptune}$ and $R_{\rm p} = 1.0R_{\rm Neptune}$. {\rev Although each class of planets are simulated at increments of $u=0.05$, at each value of $u$ the families are slightly offset from one another for clarity.}  A total of 10 thermalization events may disrupt the planet, {\rev which we denote here as ``destroyed''}. {\rev Ice giants may be frequently destroyed when chaotic tidal evolution is active.}
}
\label{thermevo}
\end{figure*}

\noindent{}In order to compute the new mode ($c_{\alpha, k}$), one first must determine the new orbital period of
the $k$th iteration ($P_k$) and recompute $\sigma_{\alpha, k}$ at the $k$th iteration. The
value of $P_k$, when summed over many pericentre passages, also helps determine $\tau_{\rm chaos}$. We finally have

\begin{equation}
P_k = 2 \pi \sqrt{\frac{a_{k}^3}{G\left(M_{\ast} + M_{\rm p}  \right)}},
\label{orbeq}
\end{equation}

\[
c_{\alpha, k} = \left(c_{\alpha, k-1} +\Delta c_{\alpha} \right) \exp{\left(-i \sigma_{\alpha, k} P_k \right)}, \ \ {\rm if}  \ E_{\alpha, k} < E_{\rm max}
\]


\[
\ \ \ \ \ = \sqrt{\frac{E_{\rm resid}}{\left| E_{\rm B,0} \right|}} \sqrt{\frac{G M_{\rm p}}{R_{\rm p}^3}}, \ \ \ \ \ \ \ \ \ \ \ \ \ \ \ \ \ \ \ \ \ {\rm if}  \ E_{\alpha, k} \ge E_{\rm max}
\]

\begin{equation}
\label{Etoohigh}
\end{equation}


\noindent{}where

\begin{equation}
E_{\rm resid} = 0.001 E_{\rm bind}
\label{Eresid}
\end{equation}

\noindent{}and

\begin{equation}
E_{\rm max} = 0.1 E_{\rm bind}
\label{Emaxeq}
\end{equation}

\noindent{}such that the binding energy of the planet is 

\begin{equation}
E_{\rm bind} \approx \frac{G M_{\rm p}^2}{R_{\rm p}}  
.
\label{Ebind}
\end{equation}

In this last step, the mode energy ($E_{\alpha, k} = \sum \Delta E_{\alpha, k}$) is capped at a fraction ($=0.1$) of the planet's binding energy. Physically, this cap represents non-linear dissipation of the mode once its amplitude becomes large. This dissipation thermalizes the orbital energy absorbed by the mode, causing inward migration. When the cap is activated, the mode amplitude is reset according to Eq. 16 of \cite{vicetal2019}, but with $E_{\alpha, k}$ replaced by $E_{\rm resid}$. The choice of the coefficients in equations (\ref{Eresid}-\ref{Emaxeq}) was explored in \cite{vicetal2019} but was not found to qualitatively affect the final orbital parameters when the planet leaves the chaotic regime.

\subsection{Criterion for ending chaotic evolution}

{\rev In order to determine when the planet does leave the chaotic regime, we cannot use equation (\ref{crit}) because that equation assumes that the f-mode contains no initial energy.
Instead we use Eq. 51 of \cite{vicetal2019}. The chaotic regime ends after the $k$th pericentre passage when

\begin{equation}
1 \gtrsim \frac{3\sigma_{\alpha,k} P_k\sqrt{\Delta E_{\alpha} E_{\rm resid}}}{\left| E_{{\rm B},k} \right|}
.
\label{critend}
\end{equation}

This equation is not as strict as equation (\ref{crit}), which would prematurely truncate the chaotic evolution if it was used as both the starting and stopping condition. The duration of chaotic evolution, and the orbital parameters at which it ceases, is then dependent on 
$E_{\rm resid}$. Larger values of $E_{\rm resid}$ allow for more extensive chaotic evolution. 
}

\subsection{Phase space exploration} \label{phase}

Now we are ready to iterate our map and determine the orbital evolution. 

\subsubsection{Orbital evolution}

Figure \ref{fourcaseJup} provides four examples of orbital evolutions for $M_{\rm p} = 1.0 M_{\rm Jupiter}$, $R_{\rm p} = 1.0 R_{\rm Jupiter}$, $a_0=10$~au and $u=\left\lbrace 1.3,1.4,1.5,1.6 \right\rbrace$. In these cases, {\rev respectively, $\tau_{\rm chaos} = \left\lbrace 0.079, 0.42, 5.3, 13.4 \right\rbrace$ Myr and the final semimajor axis is just $\left\lbrace 1.9, 2.6, 3.5, 4.8 \right\rbrace$ per cent of $a_0$.}

Our choice of $a_0=10$~au is reasonable because it implies that due to giant branch mass loss, the planet {\rev once} resided at a distance of about 3-5 au on the main sequence \citep{omarov1962,hadjidemetriou1963,veretal2011,veretal2013b,doskal2016a,doskal2016b}. That distance is sufficient for a planet to have avoided tidal engulfment throughout the giant branch phases \citep{villiv2009,kunetal2011,musvil2012,adablo2013,norspi2013,valras2014,viletal2014,madetal2016,staetal2016,galetal2017,raoetal2018}.

Figure \ref{fourcaseJup} illustrates that the evolution (i) is chaotic in semimajor axis and eccentricity, (ii) can quickly create significant changes in semimajor axis, (iii) {\rev produces small changes in} eccentricity (at most by {\rev a tenth}), (iv) {\rev calibrates changes in semimajor axis and eccentricity such that $a\left(1-e\right)$ remains nearly constant}, (v) is very sensitive to $u$, and (vi) shows a secular trend of increasing $\tau_{\rm chaos}$ as $u$ is increased. Of particular interest is the value of $\tau_{\rm chaos}$ (for equation \ref{master}), as well as the final orbital parameters that will be used as initial conditions for the non-chaotic evolution described in Section 3. 

{\rev Shown in Fig. \ref{fourcaseJup} are single evolutionary pathways for a few values of $u$. However, due to the stochasticity of f-mode evolution, a very slight change in initial conditions will produce a completely different pathway. Consequently, $\tau_{\rm chaos}$ as well as the final orbital parameters could exhibit a range of values for almost the same initial conditions.

In order to explore this variation, for every set of initial conditions, we ran 5 simulations. The only difference amongst these simulations was a tiny change in their initial value of $u$ by us adding and subtracting $1 \times 10^{-7}$ and $2 \times 10^{-7}$ to the nominal value.
}


\subsubsection{Energy evolution}

The sudden changes in semimajor axes experienced by the planets are accompanied by violent increases in internal energy. These variations can fundamentally transform the planet, inflating it and potentially destroying it. However, before the mode energy increases sufficiently highly to match the disruption energy, non-linear effects dissipate the mode energy \citep{vicetal2019}. For that reason, when the mode energy reaches a certain fraction ($10 \%$) of the binding energy (equation \ref{Etoohigh}), this energy is {\rev dissipated within the planet, with the exact location determined by the details of the non-linear breaking process; one possibility is that the energy is dissipated close to the surface and efficiently radiated away \citep{wu2018}}. Then the mode amplitude is reset. The choice of this fraction was explored in \cite{vicetal2019} and its variation was shown to have little effect on the final orbital evolution.

Hence, 10 thermalization events {\rev (assuming no energy is radiated away)} would deposit enough energy in the planet's interior to substantially alter its structure. Whether the planet would slowly inflate or be disrupted is unclear, though the former would increase the tidal dissipation rate, perhaps pushing it towards disruption. Regardless, the implications for the origin of white dwarf pollution could be important. We therefore plot the evolution of the mode energy for the planets in Fig. \ref{energyevo}, and mark with a horizontal purple line where thermalization events would occur. More thermalization events occur as $u$ is decreased: {\rev for $u=\left\lbrace 1.6, 1.5, 1.4, 1.3 \right\rbrace$, respectively, exo-Jupiters experience 7, 5, 4 and 2 thermalization events. Exo-Neptunes, at those same $u$ values, nearly all experience at least 10 thermalization events.} 

\subsubsection{Phase space exploration}

Now we can explore how $\tau_{\rm chaos}$ varies across the entire phase space of $a_0$, $M_{\rm p}$ and $\rho_{\rm p}$ as a function of $u$, when applicable. There are three limits to applicability: (i) when the planet self-disrupts, (ii) when chaotic evolution does not activate in the first place, and (iii) when chaotic evolution does not end within a computationally feasible time. These three restrictions constrain the range of $u$ which needs exploring to $u=1.10-2.00$: the incidence of thermalization increases for decreasing $u$ and non-activation of the chaotic regime occurs for high $u$. 

We simulate $u$ in increments of 0.05, {\rev and, as previously mentioned, we perform an ensemble of simulations for each set of initial conditions by varying $u$ from these nominal values by $10^{-7}$. Further, in Figs. \ref{thermevo}-\ref{chphase2}, we display results for different families of planets by applying an offset in $u$ of 0.01 to prevent overcrowding of data points. 
}

We present our results in two cases: by (i) varying $a_0$ in the exo-Jupiter case (Figs. \ref{thermevo} and \ref{chphase1}), and (ii) varying the physical properties of the planet for $a_0 = 10$ au (Figs. \ref{thermevo} and \ref{chphase1}).  

In the first case, we sampled $a_0 = 5, 10$ and 30 au. An initial semimajor axis of 5 au effectively provides a lower limit to the distance at which a giant planet that survives the giant branch phases of evolution would be planted. An initial semimajor axis of 30 au corresponds to furthest distance to which a exo-Saturn analogue would be pushed out during the giant branch phases of stellar evolution\footnote{Although scattering may occur at larger distances, computations -- even for an iterative map -- become onerous at these locations due to the extremely high eccentricity of an orbit which reaches the vicinity of the white dwarf Roche radius.}. 

In the second case, we sampled three types of extreme planets which we label as ``Light Gas Giant'' ($M_{\rm p} = 0.3 M_{\rm Jupiter}$ and $R_{\rm p} = 1.0R_{\rm Jupiter}$), ``Heavy Gas Giant'' ($M_{\rm p} = 13 M_{\rm Jupiter}$ and $R_{\rm p} = 1.0R_{\rm Jupiter}$), and ``Ice Giant'' ($M_{\rm p} = 1.0 M_{\rm Neptune}$ and $R_{\rm p} = 1.0R_{\rm Neptune}$).

First we consider the number of thermalization events in Fig. \ref{thermevo}. {\rev The figure displays a strong correlation between the number of these events and $u$.} This figure also illustrates that the number of thermalization events suffered is nearly independent of $a_0$, but has a strong dependence on basic physical structure quantities like mass and density.

Next we consider the criterion for chaotic evolution to be activated in the first place (equation \ref{crit}). In no case was chaotic evolution active for $u \ge 2.00$. As our computational limit, we adopted $10^7$ pericentre passages: all simulations exceeding this threshold were terminated due to memory and timescale considerations, as well as available resources.

Figures \ref{chphase1} and \ref{chphase2} plot $\tau_{\rm chaos}$, as well as the final values of $a$ and $u$. {\rev Plotted on the figures are the results of every simulation for which chaotic evolution is initiated and ends before $10^7$ pericentre passages and during which the planet survives.}  Both figures show similar outcomes, which itself is important and helpful. 

{\rev Notably, a spread in outcomes due to $10^{-7}$-level changes in initial $u$ manifest only on the top plots, producing a $\sim 1$ order-of-magnitude spread in $\tau_{\rm chaos}$. Further,} $\tau_{\rm chaos}$ increases with respect to $u$ in a rough power-law fashion. The final semimajor axes at the end of the chaotic regime have a single well-determined power-law correlation with {\rev initial} $u$; the translational differences in the curves are attributed to the Roche radius being a function of $\rho_{\rm p}$. Finally and importantly, in all cases {\rev changes in $u$ throughout the chaotic evolution are small but not negligible. Chaotic evolution always increases $u$, and will never push the orbital pericentre within the white dwarf Roche radius.
}

\subsection{Analytic estimation of $\tau_{\rm chaos}$}

{\rev Despite the fast speed of the iterative map to yield a result for $\tau_{\rm chaos}$ (as opposed to, for example, solving differential equations for dynamical tides), a single
explicit formula would be even faster.  Equation 53 of \cite{vicetal2019} provides the following estimate

\begin{equation}
\tau_{\rm chaos, ana} = \frac{P_0 \left| E_{{\rm B}, 0} \right| }{\Delta E_{\alpha}},
\label{chaosana}
\end{equation}

\noindent{}where $\Delta E_{\alpha}$ is assumed to be constant. Therefore, application of this formula requires one to choose $\Delta E_{\alpha}$ at a particular time.
A convenient choice would be during the first pericentre passage, in order to minimise computation.

For each one of our simulations, we computed $\tau_{\rm chaos, ana}$ and compared that value to $\tau_{\rm chaos}$. Fig. \ref{anal} displays this comparison for all of our simulations, and shows that in almost every case, $\tau_{\rm chaos, ana}$ is 0-1 orders of magnitude lower than $\tau_{\rm chaos}$. Hence, $\tau_{\rm chaos, ana}$ represents a robust order-of-magnitude
estimate of $\tau_{\rm chaos}$. Equation (\ref{chaosana}) may also then be used to determine how $\tau_{\rm chaos}$ analytically scales with different parameters. However,
the functional dependencies through $\Delta E_{\alpha}$ are nontrivial, primarily because of $K_{22,0}$.
 }

\section{Non-chaotic evolution}

If a system fails to satisfy equation (\ref{crit}), {\rev or after engaging in chaotic evolution then satisfies equation (\ref{critend}), subsequently} the orbital motion should not be modelled by chaotic energy exchange between modes and the orbit. Instead, a variety of mechanisms can dominate the evolution, including gravitational equilibrium tides, gravitational dynamical tides, thermal tides and magnetic tides. The outcome will be circularization of the orbit, and the timescale for this process to occur is $\tau_{\rm non-chaos}$\footnote{Technically, we determine circularization through $\tau_{\rm non-chaos}$ according to the first instance when $e<0.01$. Neither observational \citep{vanetal2015,manetal2019} nor theoretical eccentricity constraints \citep{guretal2017,veretal2017b} on the known minor planets orbiting around or within the tidal reach of white dwarf are more accurate than about $0.01$. We also do not incorporate any additional forces in the computation, such as general relativity, which does not secularly change the eccentricity nor semimajor axis \citep{veras2014}.}.

The recent review of \cite{mathis2018} emphasizes the complexity of modelling star-planet tides, even if only one type of the above listed tides is investigated. \cite{veretal2019} outlined a procedure for computing gravitational tides between a white dwarf and a solid body, a procedure which relies on solid mechanics \citep{efroimsky2015} and expansions from \cite{bouefr2019}. \cite{veretal2019} assumed Maxwell rheologies, adopted an arbitrary frequency dependence on the quality functions, and demonstrated that the orbital evolution is generally non-monotonic and the boundary between survival and engulfment is fractal.

\begin{figure}
\includegraphics[width=8.5cm]{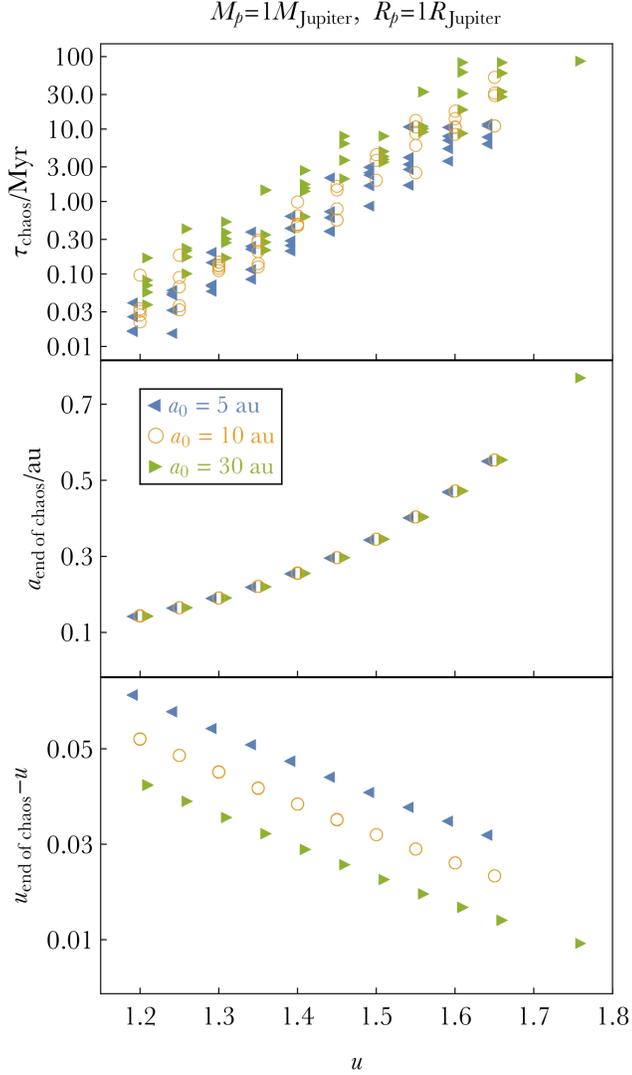}
\caption{
Values of $\tau_{\rm chaos}$ and of the final orbital parameters for different choices of $a_0$ assuming $M_{\rm p} = 1M_{\rm Jupiter}$ and $R_{\rm p} = 1R_{\rm Jupiter}$. {\rev Five different simulations were run for each pair ($u$,$a_0$) and the results are plotted only when (i) chaotic evolution ``turns on'', (ii) the planet does not self-disrupt, and (iii) the simulation was completed within $10^7$ pericentre passages.} The plots indicate that (i) the orbital pericentre must be within twice the white dwarf Roche radius in order for fast chaotic evolution to occur, {\rev (ii) for a given $u$, there is a spread in $\tau_{\rm chaos}$ but not in the final orbital parameters, (iii) the spread is confined to about one order of magnitude, and (iv) the final semimajor axis is reduced to a few to many per cent of its initial value.}
}
\label{chphase1}
\end{figure}

\begin{figure}
\includegraphics[width=8.5cm]{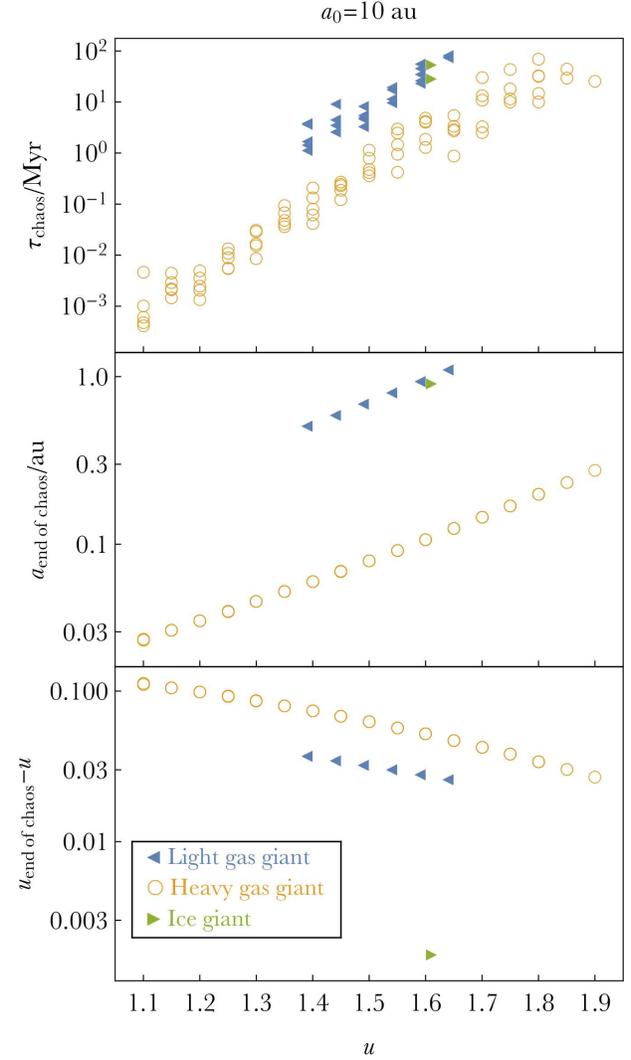}
\caption{
Like in Fig. \ref{chphase1}, with values of $\tau_{\rm chaos}$ and of the final orbital parameters, but this time for different physical planet properties, assuming $a_0 =10$~au. The three cases considered {\rev are described in the caption of Fig. \ref{thermevo}}. The plot demonstrates similar trends as in Fig. \ref{chphase1} despite the different physical properties of the planet. {\rev The dearth of green triangles arises from the fact that we have not plotted runs in which the planet may be disrupted by tidal energy deposition.}
}
\label{chphase2}
\end{figure}

Those considerations do not apply here because the planet is a gas giant and is modelled as a completely fluid body. \cite{ogilvie2014} reviewed tidal dissipation in giant planets, and emphasized again the complex way in which orbital elements are affected by different tidal components (e.g. see his Fig. 4).

Here, our objective is not to model gravitational tides in detail in the non-chaotic regime, but rather (i) to apply a simplified form to the white dwarf case, and (ii) to place non-chaotic evolution in context with $t_{\rm cool}$, $t_{\rm sca}$ and $\tau_{\rm chaos}$ (equation \ref{master}). Hence, we adopt standard treatments. We assume that the evolution is dictated by the equilibrium weak friction tidal approximation from \cite{hut1981}, where the giant planet is in a $1$:$1$ pseudosynchronous resonance with the white dwarf. The orbital semimajor axis and eccentricity then evolve according to Equations 3 and 4 of \cite{giaetal2017} as

\[
\frac{da}{dt} = \frac{9}{Q_{\rm p}'} \sqrt{\frac{G\left(M_{\ast} + M_{\rm p} \right)}{a^3}}
                            \left( \frac{M_{\ast}}{M_{\rm p}} \right) \frac{R_{\rm p}^5}{a^4} \left(1 - e^2\right)^{-15/2}
\]                             

\[
\times \left[ \frac{\left[f_2\left(e\right)\right]^2}{f_5\left(e\right)} - f_1\left(e \right) \right]
+ \frac{9}{Q_{\ast}'} \sqrt{\frac{G\left(M_{\ast} + M_{\rm p} \right)}{a^3}} \left( \frac{M_{\rm p}}{M_{\ast}} \right) \frac{R_{\ast}^5}{a^4}
\]

\[
\times \left(1 - e^2\right)^{-15/2} 
\]

\[
\times \left[
f_2\left(e\right) \left(1 - e^2\right)^{3/2} \frac{2\pi}{S_{\ast}} \sqrt{\frac{a^3}{G\left(M_{\ast} + M_{\rm p} \right)}} - f_1\left(e\right)
\right],
\]

\begin{equation}
\label{dadtequil}
\end{equation}

\[
\frac{de}{dt} = \frac{81}{2Q_{\rm p}'} \sqrt{\frac{G\left(M_{\ast} + M_{\rm p} \right)}{a^3}}
                            \left( \frac{M_{\ast}}{M_{\rm p}} \right) \frac{R_{\rm p}^5}{a^5} e \left(1 - e^2\right)^{-13/2}
\]                             

\[
\times \left[ \frac{11}{18} \frac{f_4\left(e\right) f_2\left(e\right)}{f_5\left(e\right)} - f_3\left(e \right) \right]
+ \frac{81}{2Q_{\ast}'} \sqrt{\frac{G\left(M_{\ast} + M_{\rm p} \right)}{a^3}} 
\]

\[
\times e \left(1 - e^2\right)^{-13/2} \left( \frac{M_{\rm p}}{M_{\ast}} \right) \frac{R_{\ast}^5}{a^5}
\]

\[
\times \left[
\frac{11}{18} f_4\left(e\right) \left(1 - e^2\right)^{3/2} \frac{2\pi}{S_{\ast}} \sqrt{\frac{a^3}{G\left(M_{\ast} + M_{\rm p} \right)}} - f_3\left(e\right)
\right],
\]

\begin{equation}
\label{dedtequil}
\end{equation}

\noindent{}where $Q_{\rm p}'$ and $Q_{\ast}'$ refer to the modified quality functions for the planet and star, respectively, and $S_{\ast}$ is the spin period of the star.

Each of equations (\ref{dadtequil}) and (\ref{dedtequil}) contain a component due to planetary tides and a component due to stellar tides. For main sequence planetary hosts, there are instances when both terms need to be considered. However, for white dwarfs, we can neglect the stellar tides. \cite{veretal2019} explain that the term $(R_{\ast}/a)^5$ is about 10 orders of magnitude smaller for a white dwarf than a main-sequence star, and that stellar tides through the quality function are large only when the star's viscosity is large and/or when the star spins quickly.

The neglect of the stellar tidal terms facilitate our understanding of the dependencies in the equations. In reality, $Q_{\rm p}'$ is a frequency- and time-dependent function. When considered to be constant, it just represents a scaling for the evolution. We can at least place bounds by considering several values within the extreme limits of $10^3$ and $10^7$ \citep{wu2005,matetal2010,ogilvie2014}. Further, a range of circularization timescales can then estimated if time and frequency variations are bounded between any two values within those limits, and no interdependence between the evolution of $Q_{\rm p}'$ and the orbit is assumed.


In order to provide example evolutionary sequences arising from equations (\ref{dadtequil}-\ref{dedtequil}), we continue in Fig. \ref{nonchaos1} the evolution of the $u=1.6$ curve from Fig. \ref{fourcaseJup} for five different values of $Q_{\rm p}'$. Note that the curves are self-similar, confirming that when constant, $Q_{\rm p}'$ represents just a scaling. The evolution of both the semimajor axis and eccentricity in Fig. \ref{nonchaos1} are monotonic (unlike in Fig. \ref{fourcaseJup}) and the eccentricity changes appreciably (also unlike in Fig. \ref{fourcaseJup}). 

\begin{figure}
\includegraphics[width=8cm]{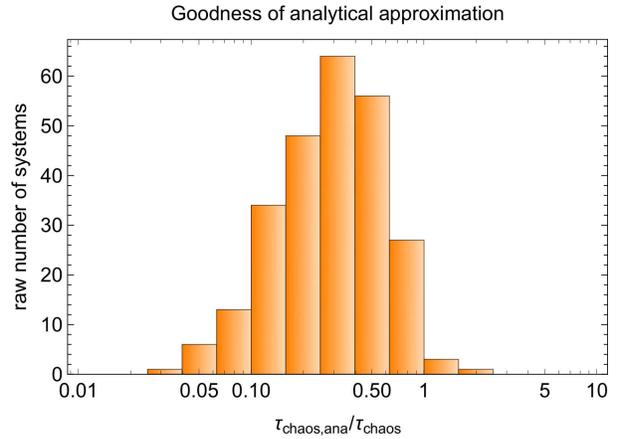}
\caption{
{\rev Comparison of the value of $\tau_{\rm chaos}$ with the simple analytical approximation from
equation (\ref{chaosana}) for every simulation for which a value of $\tau_{\rm chaos}$ was
obtained. The histogram illustrates that the analytical approximation reproduces the true
value of $\tau_{\rm chaos}$ to within about one order of magnitude. The system with the highest value on the $x$-axis is the one Heavy Gas Giant case with the large initial pericentre corresponding to $u=1.9$.
}
}
\label{anal}
\end{figure}

\begin{figure}
\includegraphics[width=8.5cm]{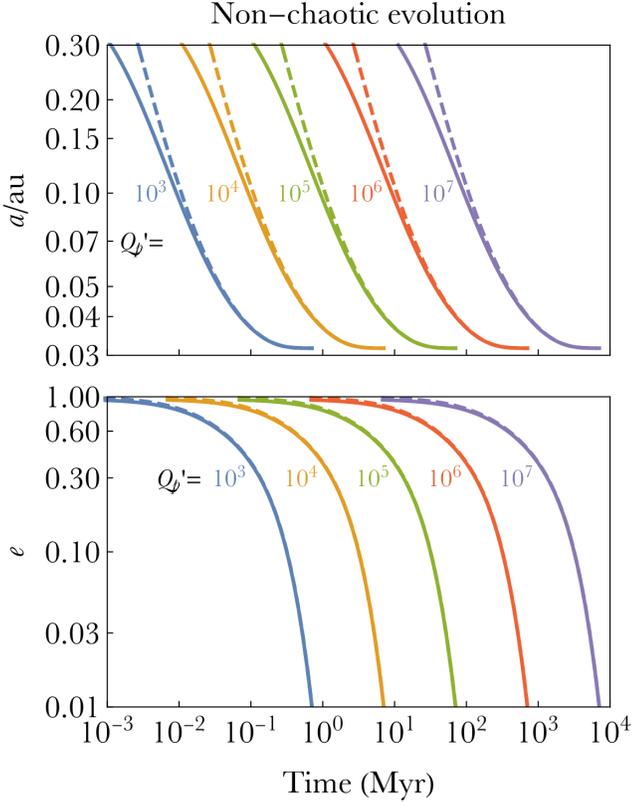}
\caption{
Continuation of the evolution of the $u=1.6$ case from Fig. \ref{fourcaseJup} in the non-chaotic regime. Different constant $Q_{\rm p}'$ values yield different potential evolutions, and hence values of $\tau_{\rm non-chaos}$; ranges of $\tau_{\rm non-chaos}$ may be estimated for time-varying values of $Q_{\rm p}'$ which are bounded between two of the curves on the plot if $Q_{\rm p}'$ is considered to be independent of $a$ and $e$. This non-chaotic orbital regime shrinks and circularizes the orbit to just outside of the Roche radius of the white dwarf. The dashed curves represent the evolution if the planet did not experience chaotic tides, but rather just equilibrium tides assuming $a_0 = 10$ au and $u=1.6$.  
}
\label{nonchaos1}
\end{figure}

Exploring the functional dependencies of $\tau_{\rm non-chaos}$ on different input parameters led us to the following empirical formula

\[
\tau_{\rm non-chaos} \approx 
\]

\[
\left(37.4 \ {\rm Myr}\right) u^{13/2} 
                                    \left( \frac{Q_{\rm p}'}{10^6} \right) 
                                     \left( \frac{M_{\rm p}}{M_{\rm Jupiter}} \right)^{-2/3}
                                    \left( \frac{\rho_{\rm p}}{1 \ {\rm g/cm}^3} \right)^{-1/2}
\]

\begin{equation}
\label{empirical}
\end{equation}

\noindent{}which is accurate to within a few per cent for the entire range of plausible phase
space for a giant planet on a highly eccentric orbit around a $0.6M_{\odot}$ white dwarf.

Equation (\ref{empirical}) is particularly useful because it allows us to avoid numerical integrations,
reveals that the dependence on $a_0$ at the start of the non-chaotic regime is weak enough 
not to be included explicitly (except through $u$), and
allows us to place limits. Crucially, the independence of $\tau_{\rm non-chaos}$ on 
$a_0$ at the start of the non-chaotic regime coupled with the {\rev small changes in} $u$ suggests that the level of decrease
of $a_0$ during the chaotic regime is not relevant for the final circularization timescale\footnote{The value of $u$ does change enough in the Heavy Gas Giant case with {\rev small} $u$ (see Fig. \ref{chphase2}) to non-negligibly shorten the circularization timescale.}.


\section{Discussion}

In this section we take stock of our results, particularly with respect to equation (\ref{master}), and discuss other relevant considerations.

\subsection{Meaning of results}

Some conclusions of our study are that chaotic mode-driven orbital evolution in white dwarf systems is particularly sensitive to $u$, occurs only when $u \lesssim 2$, and {\rev yields a value of $\tau_{\rm chaos}$ which is linked to $u$ and showcases a spread of about one order of magnitude for a given $u$}. Other conclusions are that the resulting change in $u$ is negligible and the resulting change in $a$ is significant. However, neither of these parameter significantly shifts the non-chaotic equilibrium circularization timescale through equation (\ref{empirical}). Further, $\tau_{\rm chaos}$ is largely independent of the mass, density and radius of the giant planets, whereas these variables can change $\tau_{\rm non-chaos}$  by many orders of magnitude. Consequently, the chaotic and non-chaotic regimes can be treated almost independently, which aides modelling efforts.

For a given planet discovered around a white dwarf with age $t_{\rm cool}$, if $u \gtrsim 2$ and chaotic evolution never ``turns on'', then $Q_{\rm p}'$ must be small enough to offset the high power-law dependence of $u^{13/2}$. Alternatively, for $u \lesssim 2$, both $\tau_{\rm chaos}$ and $\tau_{\rm non-chaos}$ must be considered and summed; {\rev either could be the longer timescale, especially when considering the spread in $\tau_{\rm chaos}$.}

Depending on when a white dwarf with a giant planet is observed, we can establish coupled constraints on $u$, the non-chaotic dissipation mechanisms (through $Q_{\rm p}'$, or due to a more sophisticated approach), and the time at which gravitational scattering occurs ($t_{\rm sca}$).  We can place the most stringent constraints on dissipation and orbital history for young white dwarfs. For example, a value of $t_{\rm cool}$ on the order of 10 Myr implies that separately $t_{\rm sca} < 10$ Myr and $\tau_{\rm non-chaos} < 10$ Myr. Scattering events occurring on such short timescales after the white dwarf is born has been theorized through full-lifetime numerical simulations of single-star systems \citep{veretal2013a,musetal2014,vergae2015,veretal2016,musetal2018,veretal2018} but does not yet have observational affirmation. Further, the constraint $\tau_{\rm non-chaos} < 10$ Myr usefully bounds the value of $Q_{\rm p}'$, particularly if $M_{\rm p}$ and $\rho_{\rm p}$ can be estimated.

Alternatively, giant planet detections around white dwarfs with $t_{\rm cool} \sim$~1~Gyr will not constrain tidal mechanisms and orbital history nearly as well, but still would be very useful in other manners. For example, one can place limits on the mass of planetary debris ingested in the convection zone of a metal-polluted DB white dwarf over the last Myr or so \citep{faretal2010,giretal2012,xujur2012}. These limits can range in mass over eight orders of magnitude from about the mass of about Phobos to that of Europa (see Fig. 6 of \citealt*{veras2016a}). If a giant planet is found around such a metal-polluted white dwarf with $t_{\rm cool} \gtrsim$ 1 Gyr, then that discovery would help constrain the timescales and potentially architectures of dynamical interactions between major and minor planets in that system.

\subsection{A new source of white dwarf pollution}

As suggested in the Introduction, white dwarf pollution is assumed to primarily arise from the destruction of minor planets. Major planets are generally disfavoured as the most prominently observed direct polluting source because of their small number (less than 10 per system in all known systems) and because metal sinking timescales in white dwarf atmospheres are much shorter than their cooling ages \citep{koester2009,deaetal2013,wyaetal2014,wacetal2017,baubil2018,baubil2019}. 

Nevertheless, a planet entering the Roche radius of a white dwarf will be disrupted, and some of this material may linger and pollute the white dwarf at later times. The mechanics of this process has yet to be modelled in detail. In this study, we propose that another type of disruption may act in concert: disruption created by thermal destabilization just outside of the Roche radius. This outcome is most likely for exo-Neptunes --- which are incidentally easier to scatter close to the white dwarf than exo-Jupiters --- and for small $u$. Differences in the processes of thermal disruption and gravitational disruption may have consequences for white dwarf pollution depending on how and where the planetary material is dispersed for each mechanism.

Further, although most metal pollution is generated from dry progenitors, there are striking exceptions. The pollutants in some atmospheres are volatile-rich or specifically O-rich, leading to the conclusion that the progenitors retained a substantial mass fraction of water \citep{faretal2013,radetal2015} or arose from an exo-Kuiper belt \citep{xuetal2017}.  A potential alternative explanation for the O-rich metal-polluted white dwarfs is the disruption of ice giants due to thermal destabilization.

\subsection{Comparison to main-sequence planetary systems}

The dynamical histories and tidal dissipation mechanisms of observed hot and warm Jupiters around main sequence stars are typically not as well constrained. Even for the relatively small number of host stars with accurately-measured ages (perhaps through asteroseismology), the giant planets could have migrated through their parent protoplanetary discs to their current locations rather than or in addition to being scattered there.

Metal-polluted white dwarfs contain observed circumstellar discs too \citep{farihi2016}, but these are asteroidal \citep{graetal1990,jura2003} or moon-generated \citep{payetal2016,payetal2017} debris discs whose outer radius corresponds with $u \approx 1$ \citep{gaeetal2006,manetal2016,cauetal2018,denetal2018} and are too light to have any effect on a giant planet. Further, the giant planet could not have been born in these discs \citep{perets2011,schdre2014,voletal2014,hogetal2018,vanetal2018} and must have been scattered there from au-scale distances only after the white dwarf was born.  Hence, future detections of giant planets in short-period orbits around white dwarfs give direct constraints on high-eccentricity migration that may shed light on high-eccentricity migration processes around main-sequence stars as well.

\subsection{Additional constraints}

Even if planets survive engulfment, then at the tips of the red giant and asymptotic giant branch phases, the planet is in the greatest danger of being partially or fully evaporated \citep{livsok1984,goldstein1987,neltau1998,soker1998,villiv2007,wicetal2010,beasok2011}. Our focus here is on planets which have survived these phases. Nevertheless, if a giant planet is scattered towards a white dwarf at $t_{\rm sca} \approx 0$ yr, then the planet may be evaporated by white dwarf radiation.

However, white dwarfs initially cool quickly. By adopting the analytic luminosity prescriptions from \cite{mestel1952}, \cite{bonwya2010} and \cite{veretal2015b}, we compute that a white dwarf cools to $1.0L_{\odot}$ in just 2.6 Myr after being born. If $t_{\rm sca} \lesssim 2.6$ Myr, then a relevant and interesting exercise would be to impose a time dependence on both $M_{\rm p}$ and $\rho_{\rm p}$ when computing $\tau_{\rm chaos}$ and $\tau_{\rm non-chaos}$. Evaporation during each pericentre passage is unlikely to directly shift the pericentre location non-negligibly \citep{veretal2015c}, but rather play a larger role in changing the $a$ \citep{bouetal2012}, the time-dependent solution of equations (\ref{dadtequil}-\ref{dedtequil}), and the value of $u$ through the alteration of $R$.

By itself, a scattering event, particularly without the aid of a stellar companion, raises the question of the fate of the other planet(s) in the system which created the scattering event in the first place. If any of those planets linger at sufficiently small distances, then their subsequent gravitational perturbations can prematurely disrupt mode-dominated chaotic evolution, or more severely alter the orbit after each pericentre passage. Reservoirs of small bodies, which arguably remain the most likely sources of white dwarf metal pollution, would negligibly affect a giant planet orbit.

Finally, we note that two giant substellar objects with $M_{\rm p} < 13M_{\rm Jupiter}$ have already been discovered orbiting white dwarfs, but not of the type considered here. These objects may be planets or brown dwarfs, depending on one's definition. The first, PSR B1620-26AB, is a giant body orbiting both a white dwarf and a pulsar separated by about 0.8 au in a circumbinary fashion at a distance of about 23 au \citep{sigurdsson1993,thorsett1993,sigetal2003}. The second, WD 0806-661 b, is a giant body orbiting a white dwarf at a distance of about 2500~au \citep{luhetal2011}. Prospects for finding giant planets much closer to the white dwarf in the near future are strong with TESS, LSST \citep{lunetal2018,corkip2018} and especially the final Gaia data release \citep{peretal2014}.

 \section{Summary}

Discoveries of giant planets orbiting close to white dwarfs can constrain tidal mechanisms and dynamical histories in a manner which is not available on the main sequence. Planets which survive the giant branch phases of evolution can reach the white dwarf only through a scattering event. In this work, we modelled the post-scattering tidal interaction between a white dwarf and a giant planet by using a combination of chaotic f-mode excitation and equilibrium tides. We computed the timescales for each of these mechanisms to act (Section 2 and Section 3, including equation \ref{empirical}) and determined robust dependencies on planetary mass, planetary density, initial semimajor axis and orbital pericentre. Combined with a known white dwarf cooling age (equation \ref{master}) {\rev and an expected spread in chaotic timescale evolution (top panels of Figs. \ref{chphase1}-\ref{chphase2})}, these dependencies allow one to obtain sets of scattering times and quality dissipation functions which fit both the observations and theory.

Although chaotic excitation of f-modes plays an important role in the initial circularization and high-eccentricity migration process, chaotic mode excitation {\rev ceases when the eccentricity is still large ($e \gtrsim 0.9$). Hence,} we find that the final circularization timescales are still determined by uncertain equilibrium tidal dissipation within the planet. However, chaotic mode excitation and damping can quickly thermalize a large amount of energy within planetary interiors, greater than the binding energy of ice giant planets. Depending on their response to this rapid tidal heating, these planets may become inflated or disrupted during the migration process. {\rev We found that ice giants are particularly susceptible to self-disruption if they ever enter the chaotic tidal regime.} Future constraints from detections (or lack thereof) of white dwarf planets and metal-polluted white dwarfs can constrain the dynamics of tidal migration and disruption. In particular, the cooling age of white dwarfs with planetary companions will provide an upper limit to the high-eccentricity migration timescale.

\section*{Acknowledgements}

{\rev We thank the referee for their astute and spot-on comments, which have improved the manuscript.} This research was supported in part by the National Science Foundation under Grant No. NSF PHY-1748958 through the Kavli Institute for Theoretical Physics programme ``Better Stars, Better Planets''. DV also gratefully acknowledges the support of the STFC via an Ernest Rutherford Fellowship (grant ST/P003850/1). {\rev JF acknowledges support from an Innovator Grant from The Rose Hills Foundation and the Sloan Foundation through grant FG-2018-10515.}

\label{lastpage}
\end{document}